\documentclass[12pt]{spieman}  
\usepackage{amsmath,amsfonts,amssymb}
\usepackage{graphicx}
\usepackage{setspace}
\usepackage{mathtools}
\usepackage{tocloft}
\usepackage[symbol]{footmisc}

\usepackage{lineno}
\usepackage{xcolor}

\newcommand{\add}[1]{\textcolor{black}{#1}}

\title{Estimating the statistical uncertainty due to spatially correlated noise in interferometric images}

\author[a,b,c,d]{Takafumi Tsukui}
\author[c,d]{Satoru Iguchi}
\author[c,e]{Ikki Mitsuhashi}
\author[c,d]{Kenichi Tadaki}

\affil[a]{Research School of Astronomy and Astrophysics, Australian National University, Cotter Road, Weston Creek, ACT 2611, Australia}
\affil[b]{ARC Centre of Excellence for All Sky Astrophysics in 3 Dimensions (ASTRO 3D)}
\affil[c]{National Astronomical Observatory of Japan, National Institute of Natural Sciences, 2-21-1 Osawa, Mitaka, Tokyo, Japan.}
\affil[d]{Department of Astronomical Science, SOKENDAI (The Graduate University for Advanced
Studies) 2-21-1 Osawa, Mitaka, Tokyo, Japan.}
\affil[e]{Department of Astronomy, The University of Tokyo, 7-3-1 Hongo, Bunkyo, Tokyo 113-0033, Japan}

\cftpagenumbersoff{figure}
\cftpagenumbersoff{table} 
\begin{document} 
\maketitle

\begin{abstract}
Interferometers (e.g. ALMA and NOEMA) allow us to obtain the detailed brightness distribution of astronomical sources in 3 dimensions (R.A., Dec., frequency). However, the spatial correlation of the noise makes it difficult to evaluate the statistical uncertainty of the measured quantities and the statistical significance of the results obtained. The noise correlation properties in the interferometric image are fully characterized and easily measured by the noise autocorrelation function (ACF). We present the method for (1) estimating the statistical uncertainty due to the correlated noise in the spatially integrated flux and spectra directly, (2) simulating the correlated noise to perform a Monte Carlo simulation in image analyses, and (3) constructing the covariance matrix and chi-square \add{$\chi^2$} distribution to be used when fitting a model to \add{an} image \add{with} spatial\add{ly} correlat\add{ed noise}, based on the measured noise ACF. We demonstrate example application\add{s} to scientific data showing that \add{ignoring} noise correlation can lead to significant underestimation of statistical uncertainty of the results and false detections/interpretations.
\end{abstract}


{\noindent \footnotesize\textbf{*}Takafumi Tsukui,  \linkable{tsukuitk23@gmail.com} }

\begin{spacing}{2}   

\section{Introduction}
\label{sec:intro}  
Recent developments in large interferometers (e.g.,\,ALMA and NOEMA) have made it possible to spatially resolve the brightness distribution of \add{many more} astronomical objects. These observations have enabled us to obtain a three-dimensional (R.A., Dec., and the line-of-sight velocity) structure of the gas emission and two-dimensional images of the continuum \add{within galaxies} with high spatial resolution and sensitivity. As a result, the data allow for detailed image analysis; for example, \add{investigating} spectral features of spatially resolved regions, \add{characterizing} faint and extended structures, \add{performing} Fourier analysis of the image, etc. However, the spatial correlation of the noise in interferometric images makes it difficult to evaluate the uncertainty of the results. There has been a lack of quantitative understanding of the spatial correlation of noise and methods to evaluate the statistical uncertainty of measured quantities and the significance of scientific results, such as signal detection and image analysis. 


\add{To estimate the statistical uncertainty of integrated fluxes or spectra under correlated noise, both variance and covariance of the pixel pairs in the integrated region need to be taken into account for the uncertainty propagation. Sun et al.\cite{Sun2014-nz} proposed a method based on an approximation that the covariance of noise between pixels is proportional to the synthesized beam. However, the covariance is actually proportional to the autocorrelation of the synthesized beam\cite{Refregier1998-ui}. Also, more importantly, the method approximates the synthesized beam with a single Gaussian. In contrast, the true synthesized beam has a complex structure with a main lobe inducing short-range strong noise correlation and side lobes inducing long-range weak noise correlation. 
Such an oversimplified assumption can lead to underestimation of the uncertainty. More widely used method to estimate the statistical uncertainty of integrated fluxes is based on an intuitive interpretation that the noise can be regarded as independent across beam-sized regions as described by Alatalo et al.\cite{Alatalo2013-ix}. The noise variance in the integrated values is evaluated by scaling the noise variance of individual pixels by the number of the beam area in the integrating aperture. This method also implicitly assumes the Gaussian beam to estimate the beam area of the synthesized beam. To evaluate the statistical uncertainty of the best-fitting parameters in a model fitting to the interferometric data with the correlated noise, Davis et al.\cite{Davis2017-lu} proposed a method to construct a covariance matrix from the synthesized beam, which describes the covariance of the noise and uses it to compute the $\chi^2$ value. Again, they also assumed a simple Gaussian function for the synthesized beam.}

\add{Several studies have used workarounds to avoid the need to characterize the correlated noise using Monte Carlo methods. Harikane et al.\cite{Harikane2020-nr} measure the statistical uncertainty of the integrated flux by randomly placing identical apertures to the emission-free region and adoptting the root mean square of the summed values. Boizelle et al.\cite{Boizelle2019-wg} estimate the statistical uncertainty of the fitting parameters by Monte Carlo resampling of the best-fitting parameters: adding noise extracted from the emission-free regions to the original data and refitting models. Another common technique is to fit the model in the visibility plane to measure the source size and shape where the noise in the visibility measurements is independent. This method is particularly beneficial if the source is small compared to the resolution of the interferometer since the model needs to be simple and axisymmetric for computational efficiency. However, in many cases, analysis in the image plane is necessary (e.g. complex structures such as spiral arms, bar, clumpy structures)\cite{2014A&A...563A.136M}.}

\add{There have yet to be any attempts to evaluate the statistical uncertainty for the general measurements using interferometric images (etc., integrated flux, spectra, fitting, etc.) by fully characterizing the detailed noise correlation. Refregier and Brown\cite{Refregier1998-ui} proposed to use the noise ACF to characterize the correlated noise of the Very large array (VLA) FIRST radio survey data. They used the noise ACF to explore the effect of the spatially correlated noise in the signal of the ellipticity correlation function, which encodes the imprint of the weak lensing signal by the large-scale structure of the universe. The noise ACF fully characterizes the noise correlation properties of interferometric images and provides the covariance of noise between different pixels, which allows us to measure statistical uncertainty under the noise correlation.}

In this paper, we present \add{a} method and \add{associate} Python code to characterize the spatial correlation of noise \add{in interferometric images} by measuring the noise autocorrelation function (ACF) and evaluat\add{ing} its effect on the measured quantities and the analysis results.

This paper is organized as follows.
In Sec.~\ref{sec:method}, we present the noise correlation properties of ALMA data characterized by the autocorrelation function (ACF) and show that the noise correlation originates from the synthesized beam (dirty beam) structures, which remain even in the \add{\textsc{clean}} image and cannot be removed by any deconvolution algorithm. In Sec.~\ref{sec:result2}, we introduce methods \add{for} (1) estimating the statistical uncertainties associated with spatially integrated flux or spectra\add{;} (2) generating simulated noise maps from the measured noise ACF, which are useful to estimate the statistical significance of the result obtained by any image analysis\add{;} and (3) constructing the covariance matrix from the noise ACF which can be used in the \add{$\chi^2$} formalism of the model fitting to the observed image, with example applications to real scientific data from Ref.~\citenum{Tsukui2021-sg}.

Throughout the paper, we use the noise map from emission line cube and continuum image data taken by ALMA Band 7 (2017.1.00394.S\add{;} PI\add{:} González López, Jorge) as an example, but the method proposed by this paper can be applied to other interferometric images. A Python package for easy application of the methods described in this paper, Evaluating Statistical Significance undEr Noise CorrElation (ESSENCE), is publicly available at \url{https://github.com/takafumi291/ESSENCE}. 

\section{NOISE CHARACTERIZATION OF INTERFEROMETRIC IMAGE}

\label{sec:method}  
\subsection{The characterization of spatially correlated noise}
\label{subsec:acf}
\begin{figure}[ht]
\centering
\includegraphics[width=\textwidth]{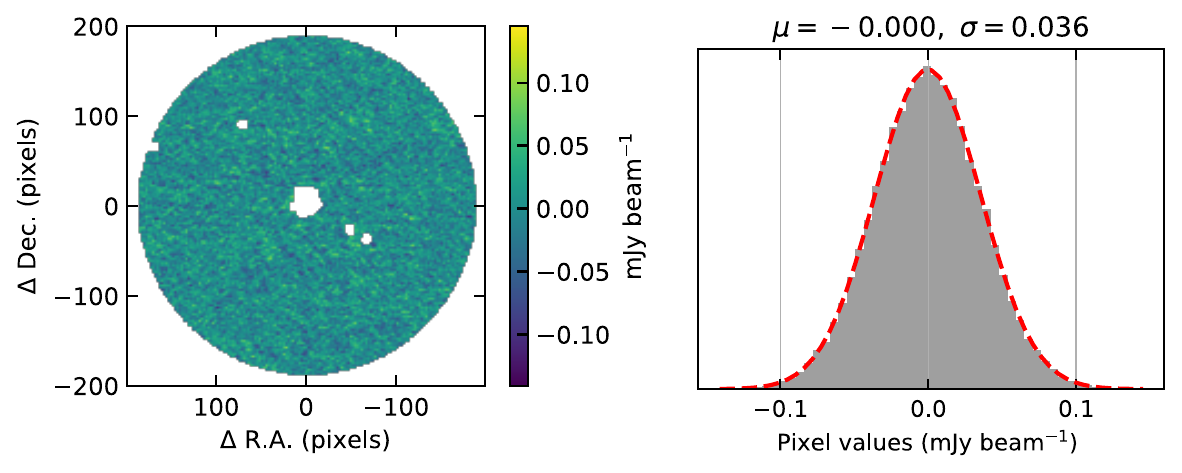}
\caption{Left: Example ALMA \add{B}and 7 noise map. The source emission region is eliminated with the 4 sigma clipping; see text. Right: The histogram of the pixel values of the noise map. The red dashed line indicates the best-fit Gaussian with the mean $\mu=0.000$ and the standard deviation $\sigma=0.036$ (mJy beam$^{-1}$). \label{fig:fig1}} 
\end{figure}

First, we consider a two-dimensional noise map $N(\mathbf{x})$, where $\mathbf{x}$ denotes the position of the pixels\add{. P}ixel regions with signal from the object of interest are excluded. The statistical properties of the noise are assumed to be uniform in the noise image, which appears to be valid in the interferometric image\footnote{We discuss this in Sec.~\ref{subsec:origin}}.  
In most of the literature, the noise in the radio interferometric image is quantified and reported with the root mean square \add{(rms)} of the noise map $N(\mathbf{x})$,
\begin{equation}\label{eq:eq1}
     \sqrt{\langle N(\mathbf{x})^2\rangle} \equiv \sigma_\mathrm{N},
\end{equation}
where the brackets denote the expected value for each pixel, which is practically estimated by averaging over the noise map. The mean of the noise in the image $\mu \equiv \langle N(x) \rangle
\approx 0$, since most of the noise represented by the system temperature $T_{\mathrm{sys}}$ is not correlated in a pair of antennas and the power of the noise does not appear in the correlator output of interferometers such as ALMA. \add{E}xtended background emission such as the cosmic microwave background (CMB) is resolved out without total power observation. However, these \add{sources of} noise contribute to the random noise associated with visibility measurements that propagate into the noise on the image by the Fourier transform. In the rest of the paper, we assume the mean of the noise map to be zero or already have been subtracted in other cases, and thus the root mean square and the standard deviation of the noise can be used interchangeably. Figure~\ref{fig:fig1} shows the example ALMA \add{B}and 7 noise map and its histogram from the observation targeting the hyper luminous infrared galaxy BRI 1335-0417 at redshift of 4.4, which will be used in the later sections. The noise map is created by eliminating \add{astronomical} source\add{s} \add{by} 4 sigma clipping \add{as well as removing} pixels \add{adjacent to these} clipped \add{regions out to} 3 \add{times the} full width of half maximum (FWHM) of the synthesized beam.
The histogram of the pixel values in the noise map is well fitted by \add{a} Gaussian \add{function}. 

When noise can be assumed to be Gaussian, the statistical and correlation properties of noise are fully quantified by the noise autocorrelation function (ACF)\cite{Refregier1998-ui},
\begin{equation}\label{eq:eq2}
\xi(\mathbf{x}_{i,j})\equiv\langle N(\mathbf{x}+\mathbf{x}_{i,j})N(\mathbf{x})\rangle,
\end{equation}
where the expected value is estimated by averaging all pairs of pixels with the relative distance $\mathbf{x}_{i,j}=(i,j)$ in the noise image. The value of the ACF noise at zero lag, $\mathbf{x}_{i,j}=\mathbf{0}$, is equal to the variance of the noise as $\xi(0)= \langle N(\mathbf{x})^2\rangle =\sigma^2_N$. When the noise has no inter-pixel correlations, the noise ACF becomes
\begin{equation}\label{eq:eq3}
       \xi(\mathbf{x}_{i,j})= 
    \begin{dcases}
    \sigma^2_N     & \text{if } \mathbf{x}_{i,j}=0\\
    0             & \text{otherwise}
    \end{dcases}
\end{equation}

To evaluate the statistical uncertainty of the derived noise ACF, we first considered the number of independent pixel pairs $N_{\mathrm{pair}}$ in the number of all available pairs $N'_{pair}$ used to evaluate the bracket in Eq.~\ref{eq:eq2}, since the pixels within a beam \add{area} are expected to be strongly correlated and not independent. We estimated the number of independent pixel pairs $N_{\mathrm{pair}}$ as the ratio of the number of all pixel pairs $N'_{\mathrm{pair}}$ and the number of pixels in the beam (beam area in pixels)\footnote{The beam area in pixels is typically estimated by $2\pi b_\mathrm{maj} b_\mathrm{min}/8\mathrm{ln}2$, where $b_\mathrm{maj}$ and $b_\mathrm{min}$ are the major and minor FWHMs of the mainlobe of the synthesized beam (\add{the ``}\add{\textsc{clean}}" beam).} $A_{\mathrm{beam}}$,
\begin{equation}\label{eq:eq4}
     N_{\mathrm{pair}}=N'_{\mathrm{pair}}/A_{\mathrm{beam}}.
\end{equation}
Then, the associated statistical uncertainty of the noise ACF $\Delta\xi(\mathbf{x}_{i,j})$ is calculated as the usual standard error of the mean but with an independent sample size $N_{\mathrm{pair}}$, that is, the standard deviation of the multiplication of the values across all pairs of pixels with separation $\mathbf{x}_{i,j}$ divided by the root of the number of independent pixel pairs $N_{\mathrm{pair}}$, 
\begin{equation}\label{eq:eq5}
     \Delta\xi(\mathbf{x}_{i,j})=\sqrt{\langle N(\mathbf{x}+\mathbf{x}_{i,j})^2 N(\mathbf{x})^2 \rangle/N_{\mathrm{pair}}}.
\end{equation}

\begin{figure}[ht]
\centering
\includegraphics[width=\textwidth]{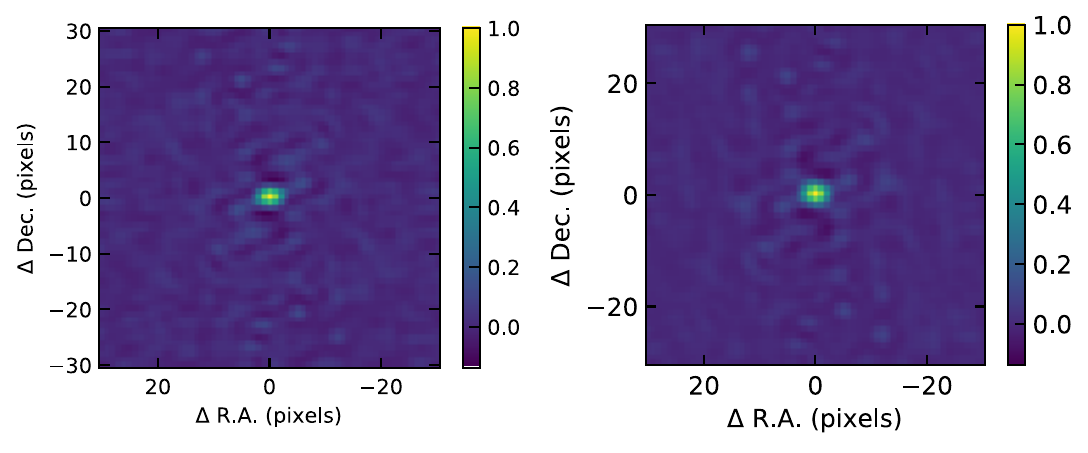}
\caption{The noise ACF computed for the ALMA Band 7 noise map (left), showing a similar pattern in the synthesized beam of the observation (right). \label{fig:fig2}} 
\end{figure}

Figure~\ref{fig:fig2} shows the results of the noise ACF (Eq.~\ref{eq:eq2}) computed for the noise map shown in Fig.~\ref{fig:fig1}, and the synthesized beam of the observation, both of which are normalized so that the central value is one. The noise ACF shows a pattern similar to that of the synthesized beam, with a correlation signal near the center and a correlation signal away from the center corresponding to the main lobe and side lobe of the synthesized beam, respectively. This suggests that most of the correlation of the noise originates from the discrete Fourier transform involved in the interferometric imaging, which is illustrated in the following subsection.

Note that the noise ACF are measured for the noise maps of the \add{B}and 7 continuum image (shown in Fig.~\ref{fig:fig2}), [C\textsc{ii}] line (velocity integrated over the velocity range of -400 to 400 km s$^{-1}$\add{, where the velocity is computed with respect to the redshifted [C\textsc{ii}] line frequency with the galaxy's redshift of 4.4074\cite{Guilloteau97}}) moment 0 map, and the [C\textsc{ii}] line cubes (each velocity channel map). These maps are primary beam uncorrected and \textsc{clean}ed images produced by the \textsc{clean} algorithm in CASA (see details in Ref.~\citenum{Tsukui2021-sg}). 

\subsection{Origin of the noise correlation}
\label{subsec:origin}
In interferometric observations, measurable quantities are visibility \add{(}Fourier amplitude and phase\add{)} of the astronomical image at the given spatial frequencies $(u,v)=\textbf{D}/\lambda$, which are related to the antenna baseline vector \textbf{D}\footnote{separation vector of pairs of antennas} projected onto the plane of the sky and the observed wavelength $\lambda$. The image is then computed by the Fourier transform of the measured visibility.

To explore the origin of the noise correlation in the image, seen in (Fig.~\ref{fig:fig2}), we start with the ideal case in which the observation measures visibilit\add{ies} at all spatial frequencies ($u,v$). The visibility of the source of interest is $V(u, v)$, which is the Fourier transform of the true flux distribution of the source in the image, $\hat{S}(x,y)=\mathrm{FT}[V(u, v)]$, where FT denotes the Fourier transform. A measurement of $V(u,v)$ usually involves uncorrelated random noise, which we describe with the random variable $\hat{N}_{\mathrm{vis}}(u,v)$ with zero mean. We assume that the statistical property of the random variable $\hat{N}_{\mathrm{vis}}(u,v)$ is uniform as a function of $u$ and $v$, that is, the system noise temperature is the same for all antennas.
The image obtained $I(x,y)$ is the Fourier transform of the measurement $V(u,v)+\hat{N}_{\mathrm{vis}}(u,v)$,
\begin{equation}
\begin{split}
    I(x,y)&=\hat{S}(x,y)+\hat{N}(x, y)\\
          &=\mathrm{FT}[V(u,v)+\hat{N}_{\mathrm{vis}}(u,v)]  
\end{split}
\end{equation}
where $\hat{N}(x, y)=\mathrm{FT}(\hat{N}_{\mathrm{vis}}(u,v))$ is the noise component of the image, which is a random variable with zero mean\footnote{\add{The} Fourier transform of the random variable with zero mean is also random variable with zero mean.}. 
The noise component of the image $\hat{N}(x,y)$ is due to the random noise associated with visibility measurements $\hat{N}_{\mathrm{vis}}(u,v)$, and the resulting noise map $N(x,y)=\hat{N}(x,y)$ is not spatially correlated in the ideal case where all spatial frequencies are measured. 

\add{In practice, visibilities are measured only at the limited spatial frequencies} \{($u_1,v_1$), ($u_2,v_2$),..., ($u_M,v_M$)\} (uv coverage). 
\add{The spatial transfer function, $W(u,v)$, is used to describe the spatial frequencies $(u,v)$ at which we measure the visibility. This function $W(u,v)$ is non-zero if the visibility at $(u,v)$ is actually measured, which can be expressed as,}
\begin{equation}
    W(u, v)=\sum_{i=0}^M \delta(u-u_i,v-v_i)+\delta(u+u_i,v+v_i),
\end{equation} 
where $\delta$ is the Dirac delta function.
The synthesized beam $b(x,y)$ is the Fourier transform of the spatial transfer function $W(u, v)$, $b(x,y)=\mathrm{FT}[W(u,v)]$. The resulting image $I(x,y)$, decomposed as the signal $S(x,y)$ from the source and noise map $N(x,y)$, is 
\begin{equation}
\begin{split}
    I(x,y) & =S(x,y)+N(x,y)\\
           & =\hat{S}(x,y)*b(x,y)+\hat{N}(x,y)*b(x,y)\\
           & =\mathrm{FT}[(V(u,v)+\hat{N}_{\mathrm{vis}}(u,v))W(u,v)],
\end{split}
\label{eq:eq8}
\end{equation}
where $*$ represents convolution.
As the noise correlation pattern (noise ACF) and the synthesized beam show a similar pattern in Fig. \ref{fig:fig2}, the noise component of the image $N(x,y)$ is the convolution product of the random variable $\hat{N}(x,y)$ and the synthesized beam $b(x,y)$. \add{Because of this,} the noise in the image \add{is well} behave\add{d}; in particular, its statistical properties are uniform over the image, as assumed in Sec.~\ref{subsec:acf}, \add{when} measur\add{ing} the noise ACF .

\begin{figure}[ht]
\centering
\includegraphics[width=\textwidth]{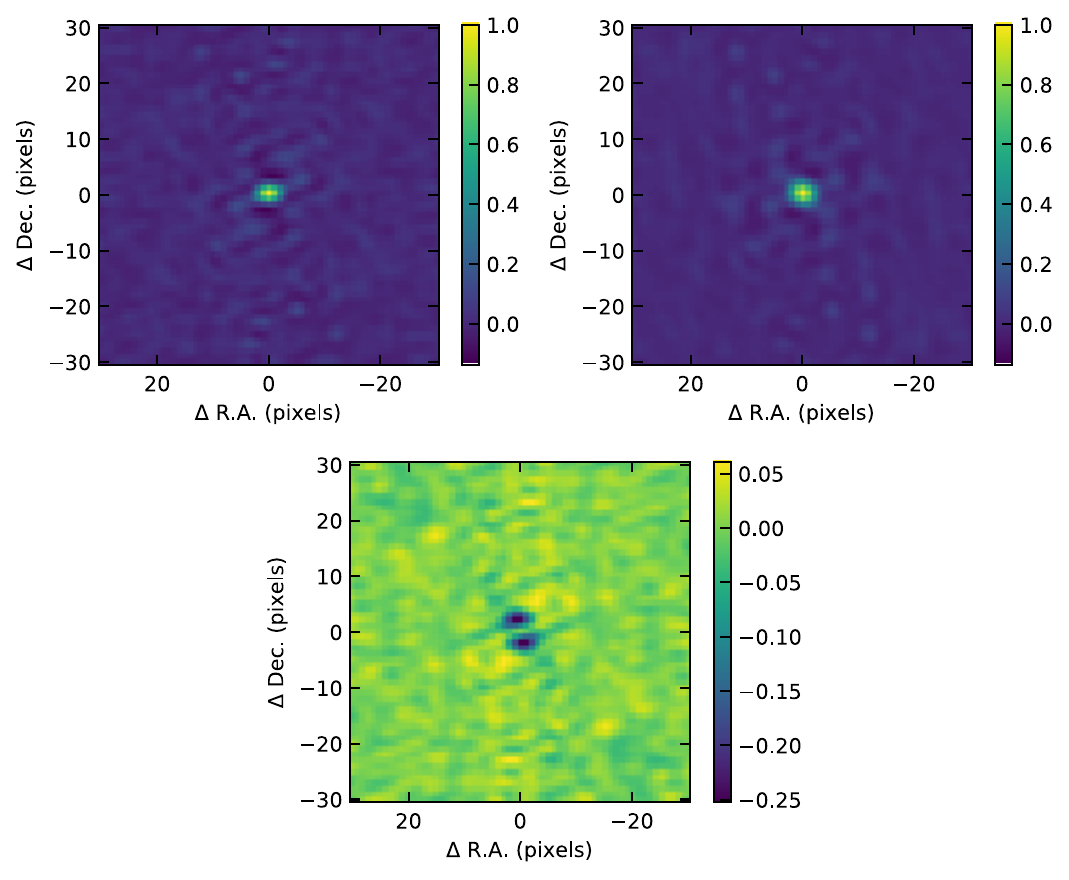}
\caption{Top left: the same noise ACF shown in Fig.~\ref{fig:fig2}, top right: the ACF of the synthesized beam, bottom: The residual of the noise ACF minus the ACF of the synthesized beam} 
\label{fig:fig4}
\end{figure}

For convenience, by replacing the sky position of \add{$(x,y)$} with the pixel position \textbf{x}, \add{the} noise map of the image in Eq.~(\ref{eq:eq8}) is written as
\begin{equation}
    N(\mathbf{x})=b(\mathbf{x})*\hat{N}(\mathbf{x})=\sum_{i,j}b(\mathbf{x}_{i,j})\hat{N}(\mathbf{x}+\mathbf{x}_{i,j}).
\end{equation}
The autocorrelation of the noise map \add{then becomes}\cite{Refregier1998-ui}
\begin{equation}
\begin{split}
        \xi(\mathbf{x}_{i,j})&=\langle N(\mathbf{x}+\mathbf{x}_{i,j})N(\mathbf{x})\rangle \\
        &=\langle \sum_{i',j'}b(\mathbf{x}_{i',j'})\hat{N}(\mathbf{x}+\mathbf{x}_{i,j}+\mathbf{x}_{i',j'})\sum_{i'',j''}b(\mathbf{x}_{i'',j''})\hat{N}(\mathbf{x}+\mathbf{x}_{i'',j''}) \rangle \\
        &=\sum_{i',j'}\sum_{i'',j''}b(\mathbf{x}_{i',j'})b(\mathbf{x}_{i'',j''})\langle\hat{N}(\mathbf{x}+\mathbf{x}_{i,j}+\mathbf{x}_{i',j'})\hat{N}(\mathbf{x}+\mathbf{x}_{i'',j''}) \rangle\\
        &=\sigma_N^2 \alpha(\mathbf{x}_{i,j}),
\end{split}\label{eq:eq10}
\end{equation}
where we used the noise ACF property of uncorrelated noise $\hat{N}$ (Eq.~\ref{eq:eq3}) for the fourth equality \add{and we have defined $\alpha(\mathbf{x}_{i,j})$ as} beam autocorrelation,
\begin{equation}
    \alpha(\mathbf{x}_{i,j}) = \sum_{i',j'}b(\mathbf{x}_{i',j'})b(\mathbf{x}_{i,j}+\mathbf{x}_{i',j'}).
\end{equation}
Equation~\ref{eq:eq10} implies that the noise ACF is related to the ACF of the synthesized beam with a constant multiplicative factor, which is the variance of the noise. In Fig.~\ref{fig:fig4} we compare the ACF of noise and that of the synthesized beam, along with the residuals (noise ACF minus beam ACF). Although the noise ACF and beam ACF show a common characteristic pattern, they do not completely coincide. \add{The difference of the two ACF shows} an extended weak positive correlation and a relatively large negative around the main beam in the residual. This disagreement is likely due to not only (1) the remaining contamination by the emission from the sources, but also (2) the process involved in the imaging of the visibility measurements. These are discussed in detail in the Appendix~\ref{sec:ap1} comparing Fig.~\ref{fig:fig4} obtained from the actual data with the one (Fig.~\ref{fig:fig12}) obtained from the simulated data with a similar observational setup and realistic noise in \add{the} visibilit\add{ies}, but without emission in the sky. \add{Note that} our interest is on the statistical property of the noise in the image plane, which we characterize by the noise ACF including effects of contamination from the source and the imaging process.

Due to the limited spatial frequency coverage, the synthesized beam $b(x,y)$ has a complex structure with sidelobes that extend from the center to a large radius. The flux from the source is spread out by the side lobes to the distant pixels in the image. The \add{\textsc{clean}} algorithm, which is most commonly used in radio imaging, deconvolves the beam pattern $b(x,y)$ for signals with high S/N ($S(x,y)/\sigma_\mathrm{N})>3$) and replaces it with a \add{\textsc{clean}} beam without side lobes (a Gaussian that approximates the mainlobe of the synthesized beam). The \add{\textsc{clean}} algorithm successfully \add{suppresses} the influence of the sidelobe and produces a high-fidelity image, but cannot remove the spatial correlations that exist in stochastic noise $N(x,y)$. Therefore, it is important to evaluate their effects on image analysis and signal detection, which we will describe in Sec.~\ref{sec:result2}. 

\section{EXAMPLE APPLICATION TO SCIENTIC DATA}
\label{sec:result2}
\subsection{Contribution of the correlated noise to the statistical uncertainty in the measured flux}

The most fundamental measurement of astronomy is the total flux \add{distributing} over some sky region in the images, which are measured by summing the pixel values over the region of interest (i.e., aperture photometry in optical astronomy). In particular, at the submillimeter band of ALMA, the flux of the continuum emission arising \add{primarily} from \add{thermal} dust, and line emission and absorption by the various atomic and molecular gases are used to estimate the physical properties of the interstellar medium (e.g., dust mass, gas mass, the energy source of the ionization or excitation, etc.). \add{I}t is important to estimate the uncertainty of the measured quantities. As shown in Fig. 2, the noise in interferometric images correlates significantly between pixels, making the estimation of the noise in the integrated flux difficult. \add{In} previous literature, statistical uncertaint\add{ies} of integrated flux\add{es} \add{were estimated} by \add{one of} two methods: (1) randomly placing \add{identical} apertures in the noise region of the image, measuring the sum within \add{each} aperture and then adopting the rms as the noise in the \add{sum of pixels in the aperture}\cite{Harikane2020-nr}; and (2) assuming that the regions in the image separated with a beam size do not correlate and adopting $\sigma_{\mathrm{N}}\add{A_{\mathrm{beam}}}\sqrt{N_\mathrm{beam}}$, where $N_\mathrm{beam}$ is the number of beams (independent regions) in the aperture\cite{Alatalo2013-ix}. $N_\mathrm{beam}$ is estimated as $A_{\mathrm{aperture}}/A_{\mathrm{beam}}$, where $A_{\mathrm{aperture}}$ and $A_{\mathrm{beam}}$ are the aperture area and the \add{\textsc{clean}} beam area \add{in pixels}, respectively. For convenience, we call methods (1) and (2) "random aperture method" and "independent beam method," respectively, in this paper. 

\add{In the "independent beam method" ($\sigma_{\mathrm{N}}A_{\mathrm{beam}}\sqrt{N_\mathrm{beam}}$), the factor $\sigma_{\mathrm{N}}A_{\mathrm{beam}}$ is the standard deviation of the sum of noise in individual pixels within a beam assuming that the noise perfectly correlates within a beam. Then the standard deviation of the sum of the noise of each independent beam area in the aperture is computed by scaling by the square root of the number of independent beams $N_\mathrm{beam}$ within the aperture. The terms $A_{\mathrm{beam}}$ and $N_\mathrm{beam}$ in the $\sigma_{\mathrm{N}}A_{\mathrm{beam}}\sqrt{N_\mathrm{beam}}$ denote just the number of data points to be summed. Therefore, we caution readers that $\sigma_{\mathrm{N}}A_{\mathrm{beam}}\sqrt{N_\mathrm{beam}}$ has the same unit as $\sigma_{\mathrm{N}}$. Most interferometric maps and measured $\sigma_{\mathrm{N}}$ are in brightness units e.g., Jy beam$^{-1}$ km s$^{-1}$ or Jy beam$^{-1}$. So we need to divide $A_{\mathrm{beam}}$ to compare with the integrated flux or spectral flux density, e.g., Jy km s$^{-1}$ or Jy. $\sigma_{\mathrm{N}}A_{\mathrm{beam}}\sqrt{N_\mathrm{beam}}$ is a factor of $A_{\mathrm{beam}}$ different from $\sigma_{\mathrm{N}}\sqrt{N_\mathrm{beam}}$ described in Alatalo et al. \cite{Alatalo2013-ix} due to the unit difference where they assume the quantity in the unit of flux.}

This section introduces how to derive the statistical uncertainty associated with the spatially integrated flux directly from the computed noise ACF. We consider adding all the pixel values at pixel positions $\mathbf{x}$ within the sky region of interest S. The random noise $N(\mathbf{x})$ in the map is characterized by the noise ACF, $\xi(\mathbf{x}_{i,j})$. The 1$\sigma$ statistical uncertainty associated with the summed value within the pixel region S, $\sigma_{\mathrm{int}}$, can be estimated as
\begin{equation}
\begin{split}
    \sigma_{\mathrm{int}}^2 & = \mathrm{Var}(\sum_{\mathbf{x}<S}N(\mathbf{x}))\\
     & = \sum_{\mathbf{x}<S}\mathrm{Var}(N(\mathbf{x}))+\sum_{\mathbf{x}<S} \sum_{\substack{ \mathbf{x}'\neq\mathbf{x}\\\mathbf{x}'<S} }\mathrm{Cov}(N(\mathbf{x}),N(\mathbf{x}'))\\
     & = N_{\mathrm{pix}}\sigma_\mathrm{N}^2+\sum_{\substack{\mathbf{x}_{i,j}=\mathbf{x}-\mathbf{x}'\\ \mathbf{x}'\neq\mathbf{x}\\\mathbf{x}, \mathbf{x}'<S} }\xi(\mathbf{x}_{i,j}),
\end{split}\label{eq:14}
\end{equation}
where $N_{\mathrm{pix}}$ is the number of pixels in the region S. Var and Cov indicate variance and covariance, respectively. The second term of the last line is the sum of the noise ACF for all possible pixel separation vectors $\mathbf{x}_{i,j}$ between two pixels within the region S. If the noise does not have an inter-pixel correlation, the second term becomes zero, resulting in $\sigma_{\mathrm{int}}=\sigma_{\mathrm{N}}\sqrt{N_{\mathrm{pix}}}$. \add{The method Sun et al. \cite{Sun2014-nz} proposed to estimate $\sigma_\mathrm{int}$ is equivalent to the Eq.~\ref{eq:14} but approximately substituting the covariance $\xi(\mathbf{x}_{i,j})$ by $\sigma_\mathrm{N}^2 b(\mathbf{x}_{i,j})$. However, $\xi(\mathbf{x}_{i,j})=\sigma_N^2\alpha(\mathbf{x}_{i,j})$ as shown in Eq.~\ref{eq:eq10}.}

\begin{figure*}[ht]
\begin{minipage}[t]{0.55\linewidth}
    \includegraphics[width=\linewidth]{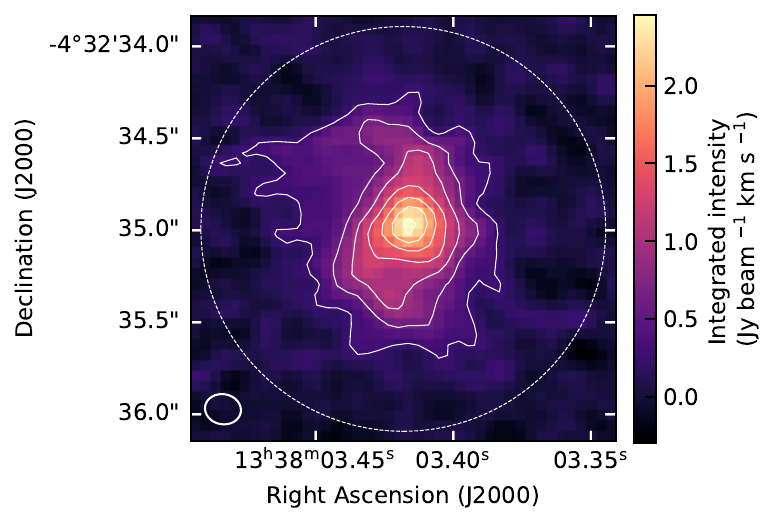}
    \caption{[C\textsc{ii}] velocity integrated intensity map of BRI 1335-0417\cite{Tsukui2021-sg} The white contour is shown every 4$\sigma$ from 3$\sigma$ to 27$\sigma$. The white elipse shown in bottom-left corner indicates the FWHM of the main lobe of the synthesized beam. The dotted line circle shows the aperture corresponding to the point with the largest number of pixels shown in Fig \ref{fig:fig6}. The [CII] spectrum within the aperture is also shown in Fig \ref{fig:fig9}. \\\label{fig:fig5}}
\end{minipage}
\hspace{0.01\textwidth}
\begin{minipage}[t]{0.45\linewidth}
    \includegraphics[width=\linewidth]{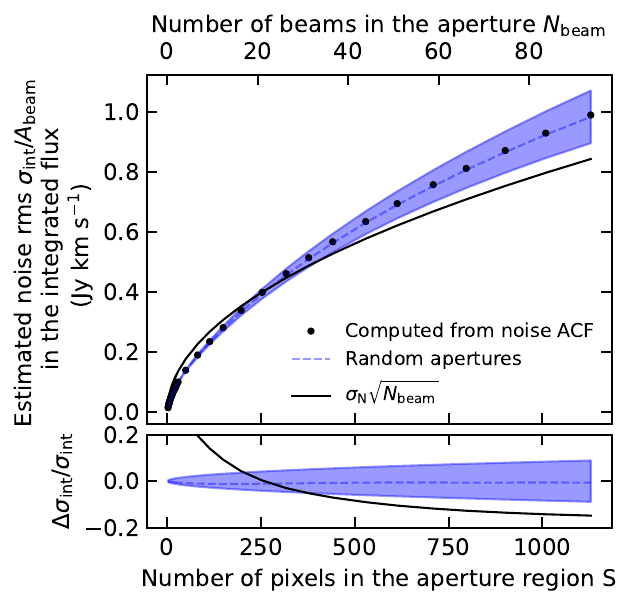}
    \caption{Top: The noise rms in the integrated flux for apertures with different sizes estimated by various methods: the one computed from the noise ACF (black points), the one estimated from the random aperture methods (blue \add{dashed} line) with the standard error (blue shade), and the one estimated from the independent beam method \add{(black solid line)}. Bottom: the fractional difference between the one computed from the noise ACF and the ones from the random aperture method and independent beam method. \\\label{fig:fig6}}
\end{minipage}
\end{figure*}

As an illustration, we used the spatially resolved [C\textsc{ii}] moment 0 map of BRI 1335-0417 taken by ALMA, shown in Fig.~\ref{fig:fig5}, where the emission spreads over multiple pixels. We calculated the noise in the integrated flux measured using a variety of apertures S with different sizes. The largest aperture is a dotted line shown in Fig.~\ref{fig:fig5}.
Figure~\ref{fig:fig6} shows the computed noise from the measured noise ACF compared to the previously used "random aperture" and "independent beam" methods. The noise calculated from the ACF is in excellent agreement with the random aperture method, while the independent beam method tends to overestimate \add{$\sigma_{\mathrm{int}}$} at smaller apertures\footnote{Considering the limiting case that the aperture is a single pixel, the obtained value by "independent beam method", $\sigma_N\add{\sqrt{A_{beam}}})$ is clearly different from the \add{fiducial} value $\sigma_N$.} and underestimate \add{$\sigma_{\mathrm{int}}$} at large apertures, showing that the assumption of "independent beam" is oversimplified.
When the field of view is small (the field of view becomes smaller at higher frequency bands in ALMA) or the aperture area becomes larger, the random aperture method cannot place apertures randomly in the limited area of the emission-free region, and thus the standard error of the estimate increases, as shown in the blue shade\add{d region} in Fig.~\ref{fig:fig6}. The proposed method \add{of calculating $\sigma_{\mathrm{int}}$}, however, can provide the best estimate by exploiting all available data to estimate the noise ACF. The total [C\textsc{ii}] flux measured with the aperture shown in Fig.~\ref{fig:fig5} is 29.51 $\pm$ 1.05 Jy km s$^{-1}$ (1$\sigma$ statistical uncertainty calculated from the noise ACF).

\begin{figure*}[ht]
\begin{minipage}[t]{0.49\linewidth}
    \includegraphics[width=\linewidth]{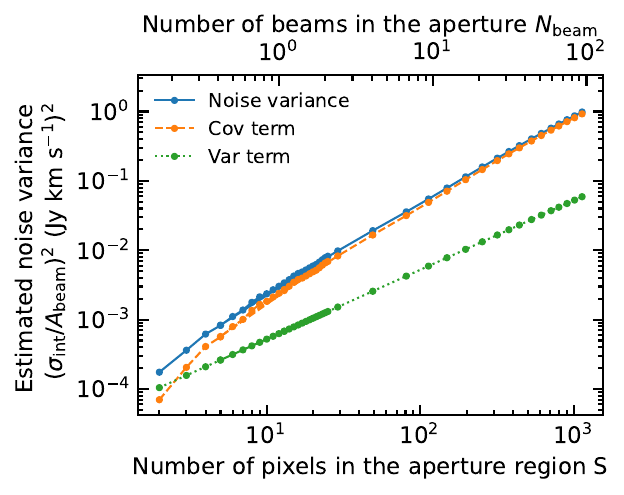}
    \caption{The measured noise variance estimated from the noise ACF for different apertures S (blue \add{points} and \add{solid} line), with the contribution from the noise variance of the individual pixels in the aperture (the first term of the last line in Eq.~\ref{eq:14}, green \add{points} and \add{dotted} line) and the contribution from the covariance due to the interpixel correlation (the second term of the last line in Eq.~\ref{eq:14}, orange \add{points} and \add{dashed} line) \label{fig:fig7}}
\end{minipage}
\hspace{0.01\textwidth}
\begin{minipage}[t]{0.49\linewidth}
    \includegraphics[width=\linewidth]{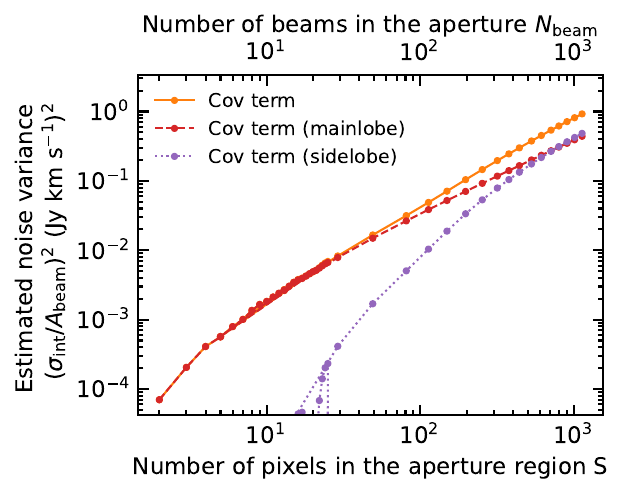}
    \caption{The covariance term due to the interpixel correlation shown in Fig.~\ref{fig:fig7}. (the second term of the last line in Eq.~\ref{eq:14}, orange \add{points} and \add{solid} line), which are further decomposed into the component of the long-range correlation due to the sidelobe of the synthesized beam \add{(purple points and dotted line)} and short-range correlation due to the mainlobe of the beam \add{(red points and dashed line)}. \label{fig:fig8}}
\end{minipage}
\end{figure*}

To demonstrate the significant effect of spatially correlated noise, Fig.~\ref{fig:fig7} shows the noise variance $\sigma_\mathrm{N}^2$ in the integrated flux calculated from the ACF along with the contributions of the noise variance of individual pixels in the aperture (the first term in Eq.~\ref{eq:14}, this is the value we obtain if we are unaware of the noise correlation) and the interpixel correlation in the aperture (the second term in Eq.~\ref{eq:14}). In the case of an aperture with two pixels, it is not mathematically allowed for the first term to exceed the second term. However, as the number of pixels in the aperture, $N_{\mathrm{pix}}$, increases, the contribution of the second term \add{dominates} over the first term because the number of the first terms is $N_{\mathrm{pix}}$ while the number of the second terms is $N_{\mathrm{pix}}(N_{\mathrm{pix}}-1)$. \add{Ignoring} the noise correlation will lead to a significant underestimation of the integrated flux uncertainty. 

Figure~\ref{fig:fig8} further divides the variance due to noise correlation into the effects of the mainlobe (correlation due to the mainlobe of the synthesized beam) and the sidelobe (long-range pixel correlation due to the sidelobe of the synthesized beam). \add{To illustrate the significance of the long-range correlation conservatively}, we define \add{short-range/long-range correlation components of the noise ACF inside/outside the ellipse with the beam FWHM in radius}, respectively. After $N_{\mathrm{pix}}$ exceeds 200, the effect of the sidelobe becomes significant. This explains the deviation of the estimate by the independent beam method from the true value (computed by the noise ACF or the random aperture method; see Fig.~\ref{fig:fig6}); the number of independent beams $N_\mathrm{beam}$ is estimated by the area of the \add{\textsc{clean}} beam, and the long-range correlation due to the sidelobe is not properly taken into account.

\begin{figure}[ht]
\centering
\includegraphics[width=0.6\textwidth]{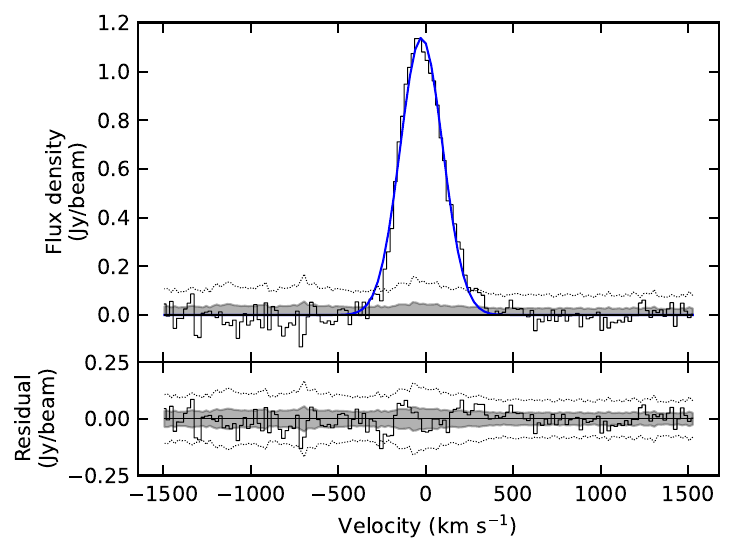}
\caption{The top panel shows the spatially integrated flux within the aperture shown in Fig.~\ref{fig:fig5} (solid black line) and the best-fit Gaussian (solid blue line). The bottom panel shows the residual (measured spectrum minus the best-fit Gaussian). The gray shade and dashed line in both panels show the statistical uncertainty of 1$\sigma$ and 3$\sigma$, respectively, calculated from the noise ACF. The figure is reproduced from Ref.~\citenum{Tsukui2021-sg} where the underlying noise is recalculated by the noise ACF of each velocity channel. \label{fig:fig9}} 
\end{figure}

Another important measurement in astronomy is the shape of the spectrum integrated over a certain region of interest. Similarly to deriving the noise in the integrated flux, we can derive the underlying noise in the spatially integrated spectrum using the noise ACFs computed for every velocity channel of the data cube. Fig\add{ure}~\ref{fig:fig9} shows the spatially integrated [CII] spectrum enclosed by the aperture shown in Fig.~\ref{fig:fig5} with 1$\sigma$ and 3$\sigma$ noise levels. The [CII] spectrum of BRI 1335-0417 is well described by a single Gaussian without deviations from the Gaussian above 3$\sigma$. Because emission lines from interesting astronomical phenomena such as outflows, tidal tails, etc.\add{,} are faint, it is important to accurately estimate the noise, otherwise it will lead to false detections.

\subsection{Simulating the noise maps}
\label{sec:simulatenoise}
In image analysis, generating random noise based on the statistical properties of the noise is useful to assess the significance of the results. In this section, we describe how to generate random noise based on the noise ACF, which fully characterizes the correlated Gaussian noise. \add{Using an example of simulated noise, we} demonstrate \add{how ignoring correlated noise leads} to misinterpretation of results. 

Once the noise ACF is measured, the noise at the \add{$\mathbf{x}_{i,j}$} positions in images with a size of $M\times M$ pixels can be generated randomly by the joint probability distribution, the probability that $N(\mathbf{x}_{i,j})$ takes the value in small intervals ($N_{i,j}+\mathrm{d}N_{i,j}$) given by\cite{Binney2008-fu}
\begin{equation}\label{eq:jointprob}
    \mathrm{d}p=\frac{\mathrm{d}N_{1,1}\cdot\cdot\cdot\mathrm{d}N_{i,j}\cdot\cdot\cdot\mathrm{d}N_{M,M}}{(2\pi)^{M^2/2}|{B}|^{1/2}}\exp{\left(-\frac{1}{2}\Sigma^M_{a,b,c,d=1}N_{a,b}{B}^{-1}_{a-c,b-d}N_{c,d}\right)},
\end{equation}
where ${B}^{-1}$ is the inverse of the matrix ${B}$ defined by
\begin{equation}
    B_{i,j}=\xi(\mathbf{x}_{i,j})
\end{equation}

\begin{figure}[ht]
\centering
\includegraphics[width=0.9\textwidth]{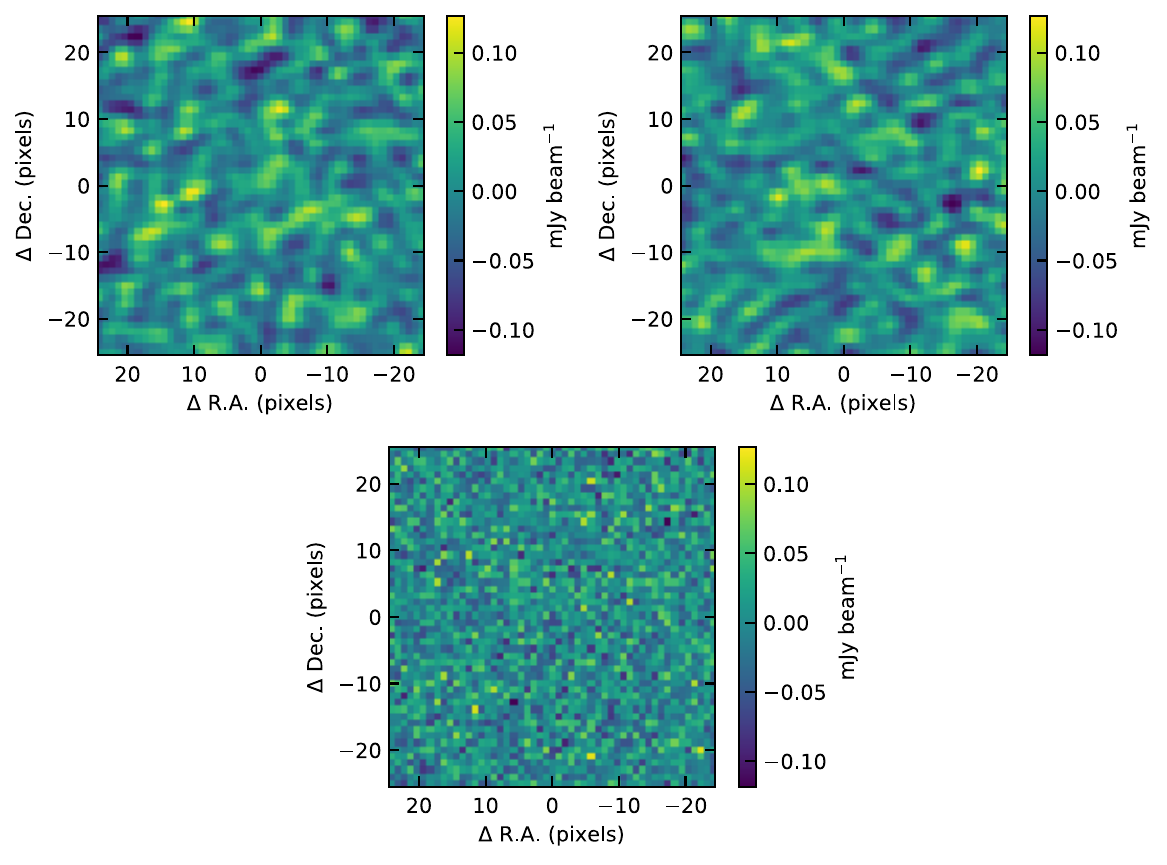}

\caption{Noise map in the observed data (top left), noise map generated from the measured noise ACF (top right), and uncorrelated noise map (bottom). All of which have the same standard deviation $\sigma_{N}$. \label{fig:fig10}} 
\end{figure}

Figure~\ref{fig:fig10} shows the comparison of the noise of the observed data, the noise randomly generated from the measured noise ACF using the joint Gaussian probability distribution (Eq.~\ref{eq:jointprob}; we used the \textsc{multivariate}\_\textsc{normal} function from the scipy package\cite{2020SciPy-NMeth})\add{,} and the spatially uncorrelated Gaussian noise, all of which have the same standard deviation \add{$\sigma_\mathrm{N}$}. The observed noise and the randomly generated noise using the noise ACF are qualitatively
similar, while the spatially correlated noise and the spatially uncorrelated noise look completely different, illustrating how dangerous it is to assume naively \add{that noise is uncorrelated} in image analysis.

In the literature, emission-free regions have been extracted and used as realistic noise maps to estimate the statistical uncertainty associated with the parameters derived by the model fitting \cite{Boizelle2019-wg}. The field of view of the interferometric observation is generally small, preventing us from obtaining a sufficient number of independent noise maps \add{or cubes} to conduct Monte Carlo experiments. \footnote{\add{As the $T_\mathrm{sys}$ is not strongly variable across a spectral window, we may use the line-free channels with cautions of the channels affected by emission lines in the atmosphere and the usually negligible variation of spatial frequencies ($u, v$) with spectral frequency.}} The proposed method can generate random noise repeatedly from the measured statistical properties of the noise (the noise ACF) in the interferometric image with the best precision limited by the available area of the emission-free region.

\begin{figure}[ht]
\centering
\includegraphics[width=0.6\textwidth]{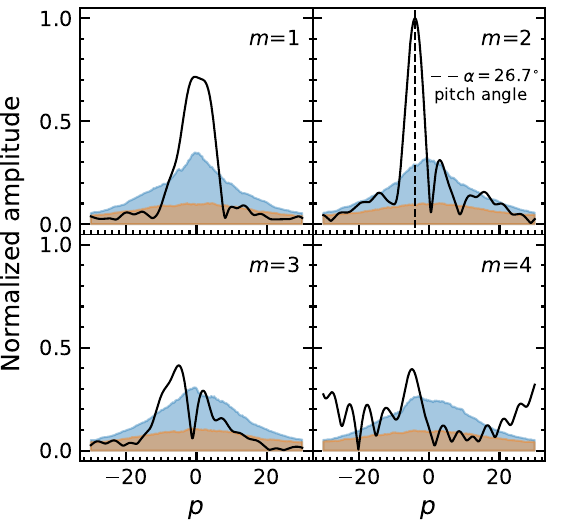}
\caption{Fourier spectra of logarithmic spiral models (solid black lines) and the underlying 1$\sigma$ noise due to the noise in the images computed from the simulated noise maps by the noise ACF (blue shaded region) and the \add{uncorrelated} Gaussian noise maps with the same standard deviation (orange shaded region). The peak in $m=2$ indicates that the dominant component of the image is a 2 armed spiral structure with a pitch angle of $\alpha=26.7$. The figure is adopted from Ref.~\citenum{Tsukui2021-sg} and the underlying noise spectrum is recalculated using the noise maps simulated by the noise ACFs. \label{fig:fig11}} 
\end{figure}

There are many potential applications of generating noise. We illustrate the significance of the correlated noise using one example: the image analysis done in Ref.~\citenum{Tsukui2021-sg}, which expands the image with series of logarithmic spirals and identifies the dominant spiral structure in the image. In Fig.~\ref{fig:fig11}, we show the Fourier spectrum of the logarithmic spirals in the [CII] intensity map of the BRI 1335-0417 shown in Fig.~\ref{fig:fig5} (black solid line) and the underlying noise spectr\add{a} estimated from the simulated noise maps from noise ACF (blue shaded region) and the simulated noise maps with no spatial correlation but the same standerd deviation $\sigma_{\mathrm{N}}$ (orange shaded region). The spectrum shows the amplitude of the logarithmic spiral with $m$ arms ($m$-fold symmetry) \add{as a function of a variable $p$, which is related to} the pitch angle of $\alpha=\arctan{(-m/p)}$\add{. To calculate the spectrum, t}he image (Fig.~\ref{fig:fig5}) was first deprojected to be viewed face-on with an inclination angle of 37.8$^{\circ}$ and a position angle of 4.5$^{\circ}$, where the package \textsc{scikit-image} performed rotation and stretching for the deprojection (which induces an additional noise correlation in the image). Then the amplitude of each Fourier component of the logarithmic spiral $m$ and $\alpha$ is calculated (see Equation~S2 in Ref.~\citenum{Tsukui2021-sg}). The noise spectrum \add{was} measured by generating 300 noise maps, calculating the amplitude of each Fourier component in the same way as the image, and taking the 84th percentiles. The noise spectrum computed by assuming the \add{uncorrelated} Gaussian noise map significantly underestimates the true noise spectrum, which could lead to the false detection of statistically insignificant structures such as the second peak in $m=2$ and multiple peaks in $m=3, 4$. The estimated noise spectra are higher than those reported in the original paper because the image stretching and rotation were not applied to the noise maps in the original paper, resulting in an underestimation of the noise level. However, the result of the original article does not change except for the updated statistical significance of each peak in $m=1,2,3,4$ being 2.1$\sigma$, 3.5$\sigma$, 1.7$\sigma$, and 1.6$\sigma$, respectively. \add{The} $m=1$ peak with 2.1$\sigma$ corresponds to the fact that the northern arm is longer than the southern arm.

\subsection{Fitting a model to an interferometric image under spatially correlated noise}\label{sec:fitting}
Fitting an analytic model of the intensity distribution to the 2D observed image is one of the most fundamental tasks in astronomy, allowing us to extract information to characterize astronomical sources from images efficiently. For a simple example, fitting 2D Gaussian to the intensity image provides essential information regarding the source's position, size, and brightness. This section introduces the construction of a covariance matrix from the measured noise ACF, which is required to calculate $\chi^2$, to properly estimate the statistical uncertainty on the parameters derived by fitting the model to the observed interferometric image.

Consider fitting a 2D model $I_{i,j}(\theta)$ to an observed image $I'_{i,j}$ with $M \times M$ pixels, where $\theta$ denotes a set of model parameters. We denote the flattened model image and the observed image by $I_{m}(\theta)$ and $I'_{m}$, respectively, with the flat indices $m=(1,2,3,...,M^2)$ that correspond to the coordinate indices $(i,j)=((1,1),...,(M, 1),(1,2),...(M,2),...,(M,M))$ in the images. The formal expression \add{to calculate} $\chi^2$ is
\begin{equation}\label{eq:cochi}
    \chi^2=\mathbf{r}(\theta)^{\mathrm{T}}C^{-1}\mathbf{r}(\theta),
\end{equation}
where $C^{-1}$ is the inverse of the covariance matrix $C$ with the size of $M^2 \times M^2$, and $\mathbf{r}(\theta)$ is the residual vector with the elements $r_{m}(\theta)=I'_m-I_m(\theta)$.
When noise has no correlation between pixels, the covariance matrix $C=\sigma_N^2E$ ($E$ is the identity matrix) and $\chi^2$ leads to
\begin{equation}\label{eq:nocochi}
\chi^2=\sum_m^{M^2}(I'_m-I_m(\theta))^2/\sigma_N^2
\end{equation}
When fitting the 2D model to the observed interferometric image, in the literature, the noise correlation has been ignored to estimate the statistical uncertainties on the derived parameters using Eq. \ref{eq:nocochi} instead of Eq. \ref{eq:cochi}; therefore, the statistical uncertainties reported are, in most cases, significantly underestimated.


We can construct the covariance matrix C in Eq. \ref{eq:cochi} from measured noise ACF by, 
\begin{equation}\label{eq:covarianceacf}
C_{m,m'}=\begin{cases}
			\xi(\mathbf{x}_{0,0})=\sigma_N^2, & \mathrm{if\ } m = m'\\
            \xi(\mathbf{x}_{i,j}-\mathbf{x}_{i',j'}), & \mathrm{if\ } m \neq m'
		 \end{cases}
\end{equation}
Here, recall that $m$ (and $m'$) are flattened indices with a one-to-one relationship with the pixel coordinate $m\rightarrow(i,j)$. As \add{$m$} and \add{$m'$} are exhangable, the covariance matrix \add{$C$} is a real-valued symmetric matrix, whose inverse can be efficiently calculated using the \textsc{linalg.pinvh} function in the Scipy package. 

\begin{figure}[ht]
\centering
\includegraphics[width=1.0\textwidth]{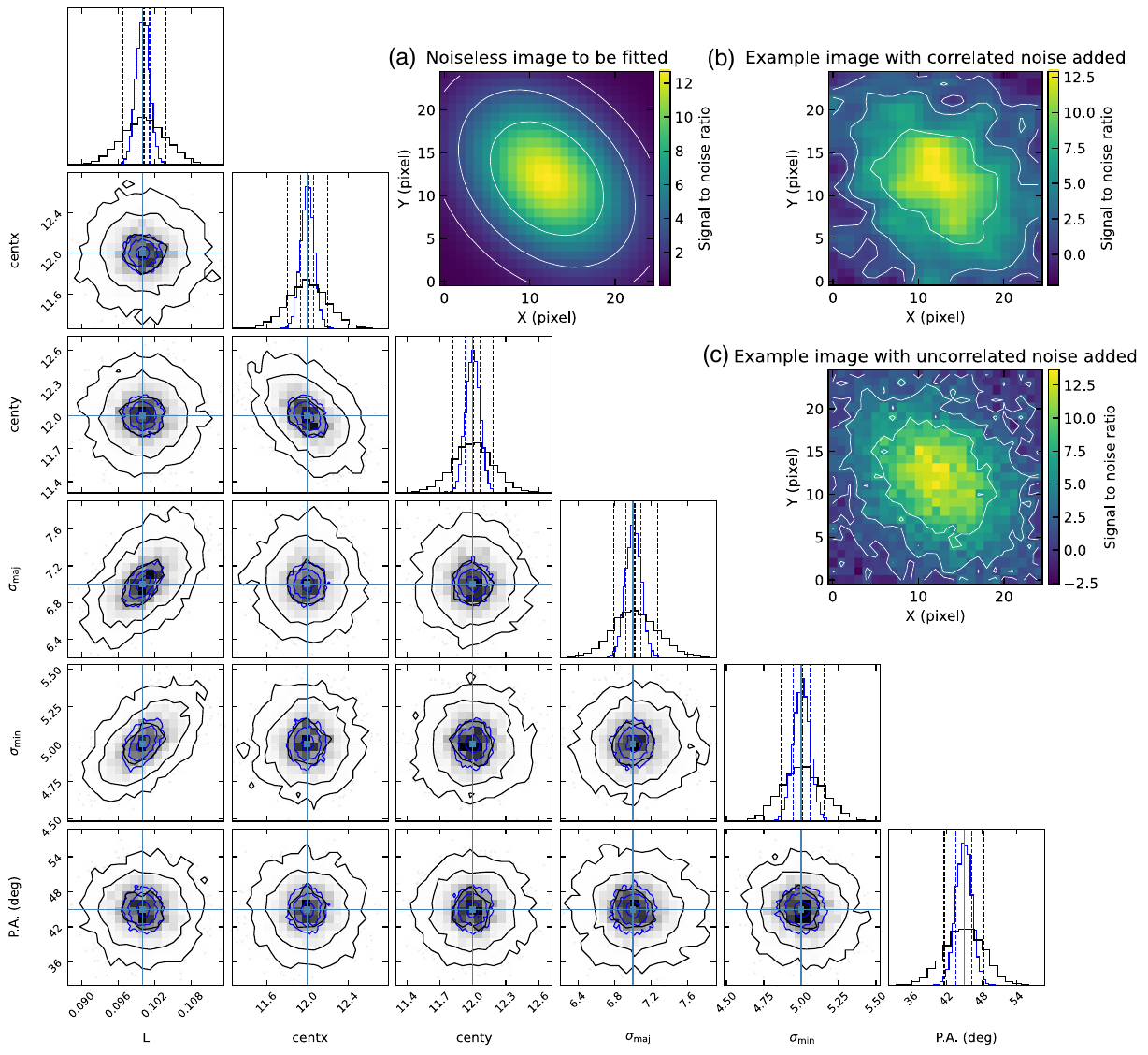}
\caption{The posterior distributions that we obtain when fitting the Gaussian model to the image with different assumptions about noise. Black: The posterior distribution sampled using the covariance matrix constructed from the noise ACF shown in Fig.~\ref{fig:fig2}. Blue: posterior distribution sampled assuming that there is no spatial correlation in the noise but with the same $\sigma_{N}$. The black distribution corresponds to the expected posterior distribution that we obtain if we properly take into account the noise correlation by measuring the noise ACF, while the blue shows that we obtain if we ignore the noise correlation and just naively use Eq.\ref{eq:nocochi} with the sky noise level $\sigma_N$. True values are shown with solid cyan lines. The three images on the top right show \add{(a)} a noiseless image to be fitted, \add{(b)} an example image with the correlated noise added, and \add{(c)} an example image with the uncorrelated noise added. \add{The comparisons with (b) and (c) illustrate that the image data with uncorrelated noise can retain more information on the intrinsic Gaussian distribution than the image with the correlated noise with the same $\sigma_{N}$.}  The white contours show emission levels of $1\sigma_N$, $3\sigma_N$, $7\sigma_N$ in both images.}
\label{fig:figex1}
\end{figure}

Now, the question \add{becomes;} what effect \add{does the} noise correlation \add{have} on the estimated statistical uncertainties of the model parameters derived from the fitting \add{process?} To explore this, we fit the 2D Gaussian model to a noiseless 2D Gaussian image and sampled the posterior distribution of the model parameters: total flux of the Gaussian distribution $L$, center of the Gaussi\add{a}n \add{(}$x, y$\add{)}, major axis and minor axis of the Gaussian \add{(}$\sigma_{\mathrm{maj}}, \sigma_{\mathrm{min}}$\add{)}, and position angle \add{(P.A.)} from the north to the major axis (counterclockwise). \add{The noise effect perturbing the best-fitting parameters is effectively taken into account by the covariance matrix $C$ in the likelihood (Eq.~\ref{eq:cochi}).} We compared the posterior distributions that resulted in two cases \add{where} (1) the noise has no spatial correlation (using Eq.~\ref{eq:nocochi}), and (2) the noise has spatial correlation characterized by the noise ACF shown in Fig.~\ref{fig:fig2} (using Eq.~\ref{eq:cochi}, and the covariance matrix constructed from the noise ACF). Note that in both cases, the noise variance in individual pixels is the same, but the latter case has spatial correlation. We used \textsc{emcee} \cite{Foreman-Mackey2013-yn} to sample posterior distributions with logarithmic likelihood $L\propto-0.5\chi^2$ and \add{a} uniform prior. Figure~\ref{fig:figex1} shows the posterior distributions of the parameters of the Gaussian model in two cases. Accounting for the spatial correlation of the noise results in a significantly larger confidence interval than that obtained assuming that there is no spatial correlation of the noise. The model parameters would appear to be constrained too well if we ignored the noise correlation in the fitting process. 

\begin{figure}[ht]
\centering
\includegraphics[width=1.0\textwidth]{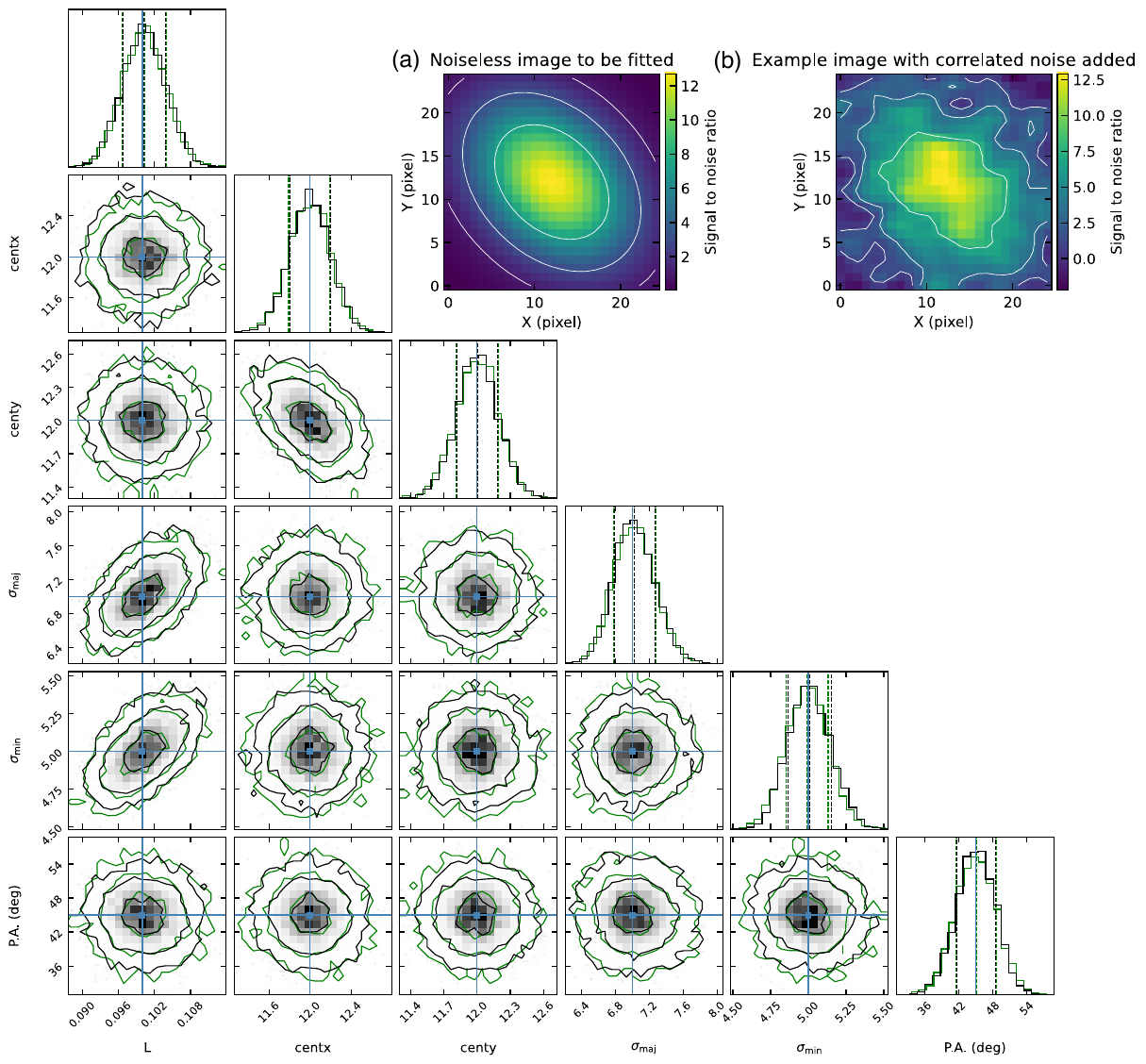}
\caption{Black: same as in Fig.\ref{fig:figex1} Green: posterior distribution estimated by Monte Carlo resampling: repeating the fitting the Gaussian model to the image added with randomly generated noise map from the noise ACF (the method described in Sec.~\ref{sec:simulatenoise}), showing the agreement between the black and green distributions. The two images on the top right\add{, (a) and (b)}, are the same as \add{(a) and (b)} in Fig.\ref{fig:figex1}.} 
\label{fig:figex2}
\end{figure}

Figure~\ref{fig:figex2} compares the posterior distribution sampled with the \add{$\chi^2$} covariance matrix (the same as Fig.~\ref{fig:figex1}) with the distribution estimated by Monte Carlo resampling\add{.} \add{The resampling is performed by} repeatedly obtain\add{ing} \add{bestfitting parameter values after adding the randomly-generated} correlated noise to the noiseless image (the method described in Sec.~\ref{sec:simulatenoise}). \add{This confirms that the posterior distributions estimated by the two approaches agree well.}

\add{In addition to the 2D fitting problem \add{when} investigat\add{ing} emission distributions, many recent studies \add{fit} 3D model cubes to observed 3D data cubes (R.A., Dec., velocity) thanks to the improvement of interferometric imaging in terms of spatial resolution and sensitivity. Radio interferometer has sufficiently high frequency resolution ($0.01$ km s$^{-1}$ or $R=3\times10^7$ at 110 GHz for ALMA), which is usually further binned to increase the signal to noise. Therefore, the noise can be regarded as independent across the frequency (or velocity) channels, and the above method described for the 2D case can be applied to compute $\chi^2$ for the 3D case. One of the main examples of 3D cube fitting is the kinematical modelling of emission lines.}

\add{There are several publicly available modelling codes for this purpose, which assume the disk geometry of the emission line, parametrize the rotation curve, velocity, dispersion, and 3D geometry of the disk (position angle, inclination), and produce 3D model cubes that can be directly compared to observations taking into account the resolution and pixel binning (e.g., \textsc{TiRiFiC}\cite{TiRiFiC07}, \textsc{KinMS}\cite{Davis2013-wm}, \textsc{3D-Barolo}\cite{Teodoro15}, \textsc{GalPaK 3D}\cite{GalPaK3D}, and \textsc{Qubefit}\cite{mneeleman_2021_4534407}). Under dynamical equilibrium, rotation curves are powerful tracers for the mass distribution of galaxies that can be further decomposed into black hole mass, stellar bulge, exponential disk, etc., depending on the data quality. These codes have been widely used to measure galaxy mass distributions and BH masses from distant to nearby galaxies (e.g., Refs.~\citenum{Davis2013-wm, Tsukui2021-sg}). However, these codes do not consider the correlation of the noise. As the number of data becomes large and the effect of noise correlation becomes significant --- estimates of the statistical uncertainty on best-fitting parameters are too small.} 

Davis et al.\cite{Davis2017-lu} estimate a covariance matrix from a point spread function and \add{use it to calculate the $\chi^2$ values.} \add{They approximate the point spread function by a single Gaussian function ignoring the side-lobe, which is an oversimplification as seen in Fig.~\ref{fig:fig8}. We find that the side-lobe effect can dominate as an image region becomes bigger. As a result, the estimates of the statistical uncertainty of parameters can be underestimated.} In addition, as seen in Sec. \ref{subsec:origin}, the point spread function does not coincide with the actual noise correlation pattern in the image. \add{Therefore, we recommend constructing the covariance matrix using Eq.~\ref{eq:covarianceacf} from the noise ACF, which fully characterizes the actual noise correlation pattern rather than from the point spread function.}

Boizelle et al.\cite{Boizelle2019-wg} \add{proposed to estimate the formal $\chi^2$ by block averaging the data and model cube to form roughly beam-sized cells to avoid the need to calculate the covariance matrix mentioned above. They stated that the block averaging method does not fully mitigate the correlation between neighbouring pixels with the presence of the long-range correlation as shown in Fig.~\ref{fig:fig8}. They estimated the final statistical uncertainty on the model parameters} by Monte Carlo realisation using line-free channels. 

\add{The example calculation for the 2D fitting} confirms that the two approaches, Monte Carlo resampling \add{(in 2D or 3D)} \cite{Boizelle2019-wg} \add{as well as} the use of the \add{c}ovariance matrix during fitting are equivalent and should provide the correct estimation under spatially correlated noise.

\section{CONCLUSIONS}
Understanding the spatial correlation of noise in interferometric images is important to correctly evaluate the statistical significance of the result. We \add{have} show\add{n} that the noise \add{autocorrelation function} \add{(}ACF\add{)} of \add{an} ALMA noise image has a pattern similar to that of the synthesized beam (dirty beam) and that the spatial correlation of the noise originates from the limited uv coverage. To correctly evaluate the statistical uncertainty of the measured quantities, we propose first measuring the noise ACF in the interferometric image, which can provide the best estimate of the full statistical properties of the noise (correlation properties) with all emission-free regions available. Once the noise ACF is measured, we can directly (1) evaluate the statistical uncertainty associated with \add{a} spatially integrated flux \add{or} spectrum, (2) randomly generate noise maps with the same correlation property, and (3) construct the covariance matrix and \add{determine a $\chi^2$} value \add{when fitting a 2D} model to an image. The method to deal with the spatially correlated noise in the interferometric image has not been documented in the astronomical literature, even for the basic spatially integrated flux \add{measurements}. We demonstrated example applications of our methods to scientific data showing that \add{ignoring} noise correlation leads to significant underestimation of statistical uncertainty of the results and false detections. A Python package for easy application of the method described in this paper, Evaluating Statistical Significance undEr Noise CorrElation (ESSENCE), is publicly available at \url{https://github.com/takafumi291/ESSENCE}. 

\add{Drizzling has become a common technique in producing optical-IR observational images. The drizzling method resamples raw under-sampled images and corrects for geometric distortion by shifting, rotating, and interpolation, and coadd these corrected images to have a common Cartesian grid. This process induces significant pixel correlation (e.g., Refs.~\citenum{Labb2003AJ....125.1107L, Sharp2015MNRAS.446.1551S}.}).
Measuring the noise ACF does not \add{require any assumption about the probability distribution of the} noise (e.g. Gaussian). Therefore, our method has potential applications to a range of astronomical images not only of interferometers but also of optical-IR observations.

\appendix 
\section{The difference between the noise ACF and the synthesized beam ACF}
\label{sec:ap1}
\begin{figure}[ht]
\centering
\includegraphics[width=\textwidth]{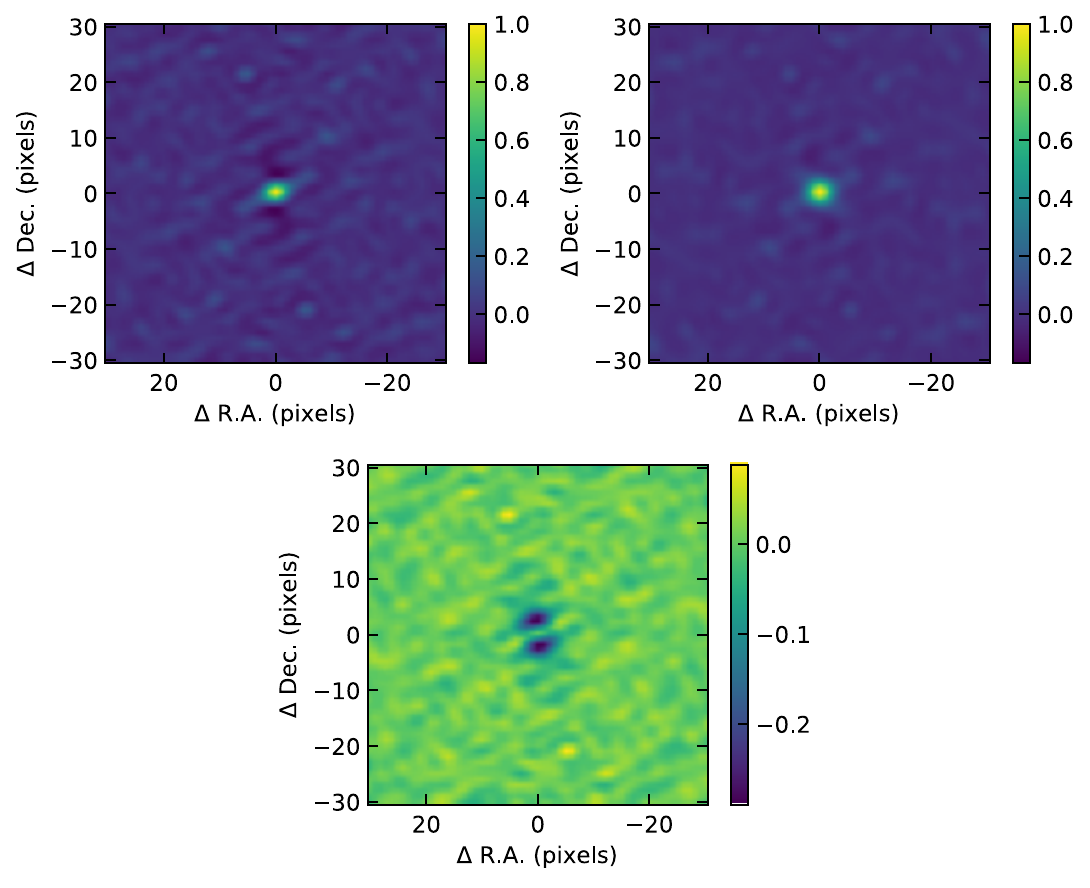}
\caption{Same as Fig.~\ref{fig:fig4} but for the noise image produced from the simulated data with a similar observational setup and without sky emission. \add{The simulated data was produced by \textsc{simalma} in \textsc{casa}.}} 
\label{fig:fig12}
\end{figure}

\begin{figure}[ht]
\centering
\includegraphics[width=0.7\textwidth]{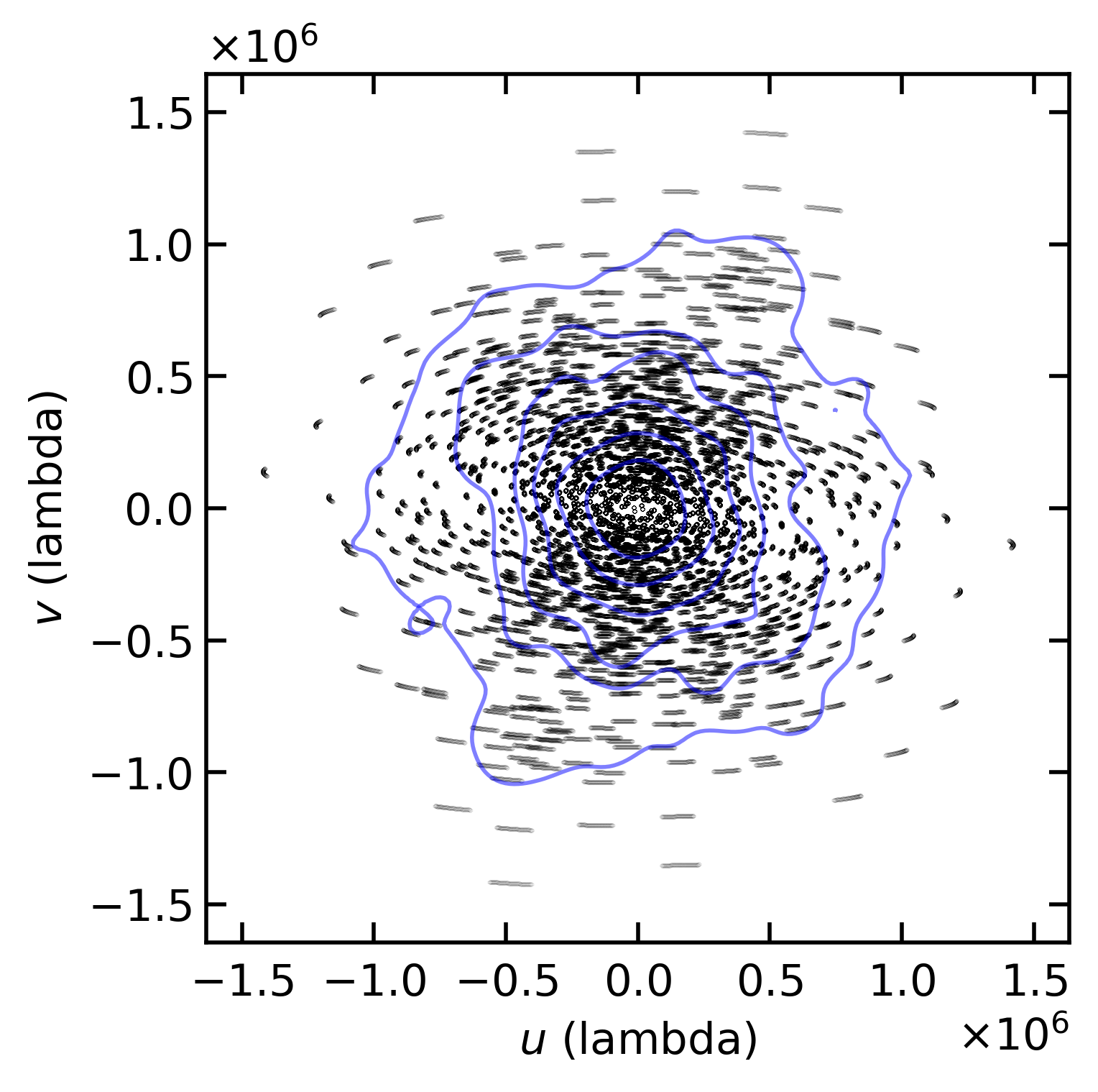}
\caption{The spatial transfer function of the simulated observation $W(u, v)$. \add{The relative density is estimated by the kernel density estimate method in the seaborn plotting package and shown in blue solid lines.}} 
\label{fig:fig13}
\end{figure}

Equation \ref{eq:eq10} implies that the noise ACF is identical to the ACF of the synthesized beam with a constant multiplicative factor. However, they do not completely coincide, showing an extended weak positive correlation and a relatively large negative around the main beam in the residual (noise ACF - ACF of the synthesized beam, Fig.~\ref{fig:fig4}). To investigate the origin of the feature, we simulate the observation with a similar setup: hour angle, sky position, antenna configuration, and realistic atomospheric noise, but without emission in the sky, using \textsc{simalma} in \textsc{casa}\cite{CASA_Team2022-cl}\footnote{The actual data is taken with $\sim$2 hour observation, with 1 hour of integration on a source intermittently separated by calibrator observations and other overheads. In comparison, the simulated observation is a continuous 1 hour of integration on the source.}. The visibility obtained is imaged with the same imaging parameters as those used for the actual data. Fig\add{ure}~\ref{fig:fig12} shows the noise ACF, the ACF of the synthesized beam, and their residual (noise ACF - ACF of the synthesized beam) for the simulated observation without sky emisssion. The residual does not show the extended positive correlation pattern seen in Fig.~\ref{fig:fig4}, suggesting that the positive correlation pattern may be due to the sky emission of the sources and undetected background sources. The emissions from the astronomical sources are spread in the noise region by long-range sidelobes of the synthesized beam, which are generally removed for bright emissions down to the noise level of 1.5 to 3 sigma by \textsc{clean}. Weaker emissions and its leakage into the surrounding pixels remain in the noise region. These un\textsc{clean}ed components and weak background sources hidden in the noise map may produce the faint positive correlation pattern in the residual Fig.~\ref{fig:fig4}.
However, the residual obtained for the simulated observation without sky emission (Fig.~\ref{fig:fig12}) still shows a similar negative around the main beam as seen in Fig.~\ref{fig:fig4}, suggesting that the large negative cannot be attributed to emissions from the sky and may be due to the process involved in the imaging. In practice, the Fourier transform in the imaging is performed by the discrete Fourier transform (FFT), in which the visibility data is evaluated discretely \add{using a} rectangular grid. The visibility data represented as ($V(u,v)+\hat{N}_{\mathrm{vis}}(u,v))W(u,v)$ in Eq.~\ref{eq:eq8} are interpolated\footnote{more precisely, the visibility data are convolved by a function to produce \add{a} continuous distribution. The function is chosen \add{to minimize} the image aliasing.\cite{Thompson2017-fp}} to estimate the visibility value in the center of the grid. Each grid is further weighted depending on the number of data points in the grid cell or the rms error of the data in the cell\cite{Briggs1995-yt}. The negative feature in the north-south direction around the main beam may correspond to the fact that the sampling of $(u,v)$ is relatively denser in the north-south direction (see Fig.~\ref{fig:fig13}), which is more affected by the averaging process, than in the east-west direction.

\subsection*{Disclosures}
There are no competing interests to declare in this paper.

\subsection*{ACKNOWLEDGEMENTS}
This paper is based on the scientific content previously reported in SPIE proceedings, Ref~\citenum{10.1117/12.2629721} with the additional section Sec.~\ref{sec:fitting} which describes how to construct the covariance matrix and the $\chi^2$ distribution from the noise ACF, which can be used to fit a model to the observed image under spatially correlated noise. We are grateful to the anonymous reviewers for giving insightful comments, which improved the manuscript significantly. TT deeply thanks \add{Emily Wisnioski and} Ma, Yik Ki (Jackie) at Australian National University, and Timothy A Davis at Cardiff University for providing constructive and insightful feedback on the manuscript. TT also thanks \add{Cameron Van Eck}, Alice Richardson, Daisuke Iono, Jorge A. Zavala, and Izumi Takuma for encouraging comments and helpful discussions. TT was supported by the ALMA Japan Research Grant of NAOJ ALMA Project, NAOJ-ALMA-264. This research has used the following data: ADS/JAO.ALMA \#2017.1.00394.S. ALMA is a partnership of ESO (representing its member states), NSF (USA) and NINS (Japan), together with NRC (Canada), MOST and ASIAA (Taiwan), and KASI (South Korea), in cooperation with the Republic of Chile. The Joint ALMA Observatory is operated by ESO, AUI/NRAO, and NAOJ. Data analysis was carried out in part on the common-use data analysis computer system at the East Asian ALMA Regional Center (EA ARC) and the Astronomy Data Center (ADC) of the National Astronomical Observatory of Japan (NAOJ). 

\subsection* {Code, Data, and Materials Availability} This research made use of \textsc{numpy}\cite{harris2020array}, \textsc{scipy}\cite{2020SciPy-NMeth}, \textsc{matplotlib}\cite{Hunter:2007}, \add{\textsc{seaborn}\cite{Waskom2021} ,} Astropy,\footnote{http://www.astropy.org} a community-developed core Python package for Astronomy \cite{astropy:2013, astropy:2018}, and \textsc{spectral-cube}\cite{2019zndo...2573901G}. A Python package of the method described in this paper, Evaluating Statistical Significance undEr Noise CorrElation (ESSENCE), is available at \url{https://github.com/takafumi291/ESSENCE}. 


\bibliography{article}   
\bibliographystyle{spiejour}   


\vspace{2ex}\noindent\textbf{Takafumi Tsukui} is an postdoctoral fellow at the Australian National University. He received his BS degree in physics from Tohoku University, and his PhD degree in physics from Graduate University for Advanced Studies, SOKENDAI. He is a member of \add{the Society of Photographic Instrumentation Engineers (}SPIE\add{)}, American Astronomical Society (AAS), and Astronomical Society of Japan (ASJ).

\vspace{2ex}\noindent\textbf{Satoru Iguchi} received his PhD in engineering from the University of Electro-Communications, Tokyo, Japan. Currently, he is a professor at the National Astronomical Observatory of Japan of the National Institutes of Natural Sciences and also a professor of radio astronomy at the Graduate University for Advanced Studies.

\vspace{2ex}\noindent\textbf{Ikki Mitsuhashi} received his BS degree in physics from Tohoku University and his master’s degree in physics from the University of Tokyo. He is a PhD student at the University of Tokyo.

\vspace{2ex}\noindent\textbf{Kenichi Tadaki}: Biography is not available.

\listoffigures
\listoftables

\end{spacing}
\end{document}